\documentclass{cta-author}
\usepackage{multirow}
\usepackage{makecell}
\usepackage{bm}
\usepackage{caption}
\usepackage{graphicx, subfigure}
\usepackage{xcolor}
\usepackage{float}

{}
{}
{}

\begin{document}

\supertitle{Brief Paper}

\title{A Fast algorithm of PET System Response Matrix based on Straight Line Truncation}

\author{\au{Zhenzhou Deng$^1$} \au{Feifan Luo$^1$} \au{Anyi Li$^2$} \au{Liang Ling$^1$}  \au{Jiu Xiong$^1$}  \au{Sikang Xu$^1$}  \au{Hongsheng Deng$^1$}  \au{Xin Zhao$^1$}}

\address{
\add{1}{School of Information Engineering, Nanchang University, Nanchang, 330031, People's Republic of China}
\add{2}{Jiluan Academy, Nanchang University, Nanchang, 330031, People's Republic of China}
\email{zzdeng@ncu.edu.cn}}

\begin{abstract}
\looseness=-1 In the calculation of Positron Emission Tomography (PET) image reconstruction, System Response Matrix (SRM) characterizes the numerical relationship between measurement space and image space. Due to a significant amount of calculation and low efficiency of the current SRM algorithm, the application of iterative reconstruction in the medical and industrial fields is limited. The time and space of computing SRM are the most important factors that researchers are concerned about. In this paper, the Straight Line Truncation (hereinafter referred to as SLT) method is applied to calculate the SRM. We verified that it improves the speed of calculating the SRM and reduces its spatial complexity. The experimental results show that the forward projection speed of SLT algorithm is six times more than that of the traditional Ray-Tracing algorithm. Besides, the forward projection speed of the small angle of SLT is about 23 times that of the traditional Ray-Tracing algorithm, and the storage space of SLT is about one sixty-fourth of the storage space of traditional Ray-Tracing algorithm.
\end{abstract}

\maketitle

\section{Introduction}\label{sec1}

PET is a nuclear medicine imaging technology, which is the latest development of current imaging technology and the combination of nuclear medicine technology and tomography technology. It is a major advance in the field of nuclear medicine and plays a crucial role in clinical diagnosis \cite{1,2}. It is a non-invasive medical imaging technique, which provides three-dimensional images reflecting the functional and biological processes. Therefore, it becomes more and more crucial in the early diagnosis, treatment planning and curative effect evaluation of cancers \cite{1,3,4}. Additionally, it also widely used in neurology \cite{5} and cardiology \cite{6}.

Iterative reconstruction is the classic reconstruction algorithm for PET. Comparing with the analytical methods, such as Filtered Back Projection (FBP), iterative reconstruction techniques can get relatively better imaging statistical performance and improve the algorithm's robustness. Besides, it also has advantages on the compatibility with incomplete data. On the other hand, iterative reconstruction has a large amount of calculation, which affects its application. In iterative reconstruction, whether it is forward projection or back projection, the projection coefficient is used repeatedly, which requires not only a large amount of storage space, but also huge calculation \cite{7}.

Many scholars have proposed different methods to improve the speed and accuracy of reconstruction and reduce the required storage space.
HERMAN G and MEYER L pointed out the selection of relaxation factors in 1993 to improve the reconstruction speed \cite{8}. GUAN H and GORDON R in 1994 optimized the speed and accuracy of Algebraic Reconstruction Technique (ART) by studying the access mode of projected data \cite{9}. In 2002, Huaxia Zhao presented an improved method based on Ray-Tracing techniques for a Field Of View (FOV) with an irregularly bounded set of voxels. It accelerates the conventional Siddon algorithm by removing redundant projection operations in regions where there is no activity \cite{10}. Huaxia Zhao also raised a new Ray-Tracing computing method in 2003, which is applied to a list-mode Expectation-Maximization (EM) reconstruction algorithm as a part of the system model to improve its projection speed \cite{11}. In 2007, Jian Zhou put forward wavelet-based Maximum A Posterior (MAP) EM algorithm for PET reconstruction, named WV-MAP-EM and it has computational advantages due to the availability of fast algorithms \cite{12}. In 2010, Yong Long proposed two separable footprint projector methods that replace the voxel footprint functions with 2D separable functions in 3D image reconstruction, so as to ease the computation burden of 3D cone-beam forward and backward projectors \cite{13}. In 2014, Wenbin Wu reduced the memory consumption and improved the computational efficiency by making use of the symmetry of modular Ray-Tracing \cite{14}. With the development of machine learning and neural networks, they are extensively used to improve the quality of reconstructed images. In 2015, Guobao Wang proposes a kernel-based method that reconstruct PET image according to a set of features obtained from prior information. It is easier to implement and provides better image quality for low-count projection data than other regularization-based methods \cite{15}. Kuang Gong trained a deep residual convolutional neural network and embed the neural network in iterative reconstruction framework to improve PET image quality \cite{16}. Besides, denoising methods \cite{17} and guided image filtering \cite{18} have been applied to optimize accuracy of reconstruction. Although there are many improved algorithms, increasing computational efficiency, reducing memory usage, and improving accuracy are still the major research objectives for PET image reconstruction.

Ray-Tracing \cite{19,20} is a classic model of SRM. It is simple to implement and suitable for occasions where resolution is not critical but real-time performance is required. Over the past several years, Ray-Tracing has been widely and successfully applied to medical imaging technology, such as Magnetic Resonance, Computed Tomography and PET \cite{21,22,23}.

Lines of Response (LORs) are symmetrical (including central symmetry and axial symmetry). Therefore, every 8 LORs can be regarded as a group, and the complete SRM can be obtained by calculating the projection coefficients of one LOR in every group. With the symmetry, the calculation order of the SRM is changed, and the amount of calculation is reduced by several times compared with the traditional calculation order \cite{24}. Based on Ray-Tracing model and the idea of symmetric block, we propose a new algorithm to calculate SRM, named SLT algorithm. Different from existing algorithms, we correct the initialized SRM instead of calculating it directly, which can greatly shorten the time. At the same time, only the non-zero value of the SRM is handled. Therefore, this method not only ensures accuracy and improves the reconstruction speed greatly, but also reduce the space complexity. Obviously, the proposed method is highly suitable for iterative reconstruction algorithm. This paper is organized as follows. Section 2 briefly describes the discretization process of Ray-Tracing model and expounds the SLT algorithm. Section 3 shows numerical experimental results. Finally discussion and conclusions are drawn in Section 4.

\section{Methods}\label{sec2}

\subsection{Discretization process of Ray-Tracing model}\label{sec2.1}
We denote the image of size $n \times n$ as
$\bm{F}=\left[ \begin{array}{cccc}
f_1, & f_2, & \cdots,& f_N \\
\end{array}
\right ]^T$ , where $N=n^2$. A discrete representation of Ray-Tracing in the absence of noise can be described by a system of linear equations:
\begin{align}\label{eq1}
&p_j=\sum_{i=1}^N a_{ji}\times f_i\notag\\
&=\left[ \begin{array}{cccc}
       a_{j1}, & a_{j2}, & \cdots,& a_{jN} \\
       \end{array}
       \right ]
    \left[ \begin{array}{cccc}
       f_1, & f_2, & \cdots,& f_N \\
       \end{array}
       \right ]^T
\end{align}
where $p_j$ is the projection data, $i$ is the pixel index, weight $a_{ji}$ is the length of intersection of the $j^{th}$ LOR with the pixel $i$, and $f_i$ is the value at pixel $i$ in the image space.
Its matrix form is expressed as
\begin{equation}\label{eq2}\bm{P_{K \times 1}}=\bm{A_{K \times N}} \times \bm{F_{N \times 1}}\end{equation}
In this expression, $K$ is the number of LORs,

$\bm{P}=\left[ \begin{array}{cccc}
p_1, & p_2, & \cdots,& p_K \\
\end{array}
\right ]^T$,

$\bm{A}=\left[ \begin{array}{cccc}
a_{11}, & a_{12}, & \cdots,& a_{1N} \\
a_{21}, & a_{22}, & \cdots,& a_{2N} \\
\cdots, & \cdots, & \cdots,& \cdots \\
a_{K1}, & a_{K2}, & \cdots,& a_{KN} \\
\end{array}
\right ]$.
\\Here, $\bm{P}$ is the measurement vector, $\bm{F}$ is the image vector and $\bm{A}$ is called the SRM \cite{25}. Generically, different modeling methods are manifested in the difference of SRM. The accuracy and precision of the SRM elements determine the quality of the reconstructed image \cite{26}. Furthermore, the time consumed in calculating SRM affects the speed of the reconstruction process. Without exception, the focus of our algorithm is also the calculation of SRM.

\subsection{SLT algorithm}\label{sec2.2}
In this part, we introduce the idea of SLT method. First, we assume that all the LORs are only divided by the lines parallel to the x-axis. Each LOR is divided into a number of line segments, which are equal in length. Corresponding to this basis, we initialize the length of line segments. But some of these line segments are also divided by the lines parallel to the y-axis. So we need to correct the length of line segments on the basis of the hypothesis reasonably. Here we present the detailed process of the algorithm, which combines SLT method with the idea of symmetric block.

In this paper, an image of size $n \times n$ is discussed, as shown in Fig. 1, where the pixel width is 1, the distance between LORs is 1, origin $O$ is at the center of the image space, and $\theta$ is the angle between the detector and the horizontal line. Based on the idea of symmetric block, we take out one LOR from each group and put them into a new group named calculation group. To ease the computation burden, we processed the calculation group. We choose LORs which simultaneously satisfy three conditions as the elements of the calculation group: (i) at 1-45 degrees, (ii) on the left side of origin $O$, (iii) intersect with the image. In the figure, only the calculated LORs are drawn, where the solid lines are the LORs intersecting with the upper edge of the image and the dotted lines are the LORs intersecting with the left edge of the image. $N_{up}$ is the number of LORs at $\theta$ degrees that intersect the upper edge of the image, and $N_{total}$ is the number of LORs at $\theta$ degrees that intersect the lower edge of the image. Obviously, the equation of the LOR closest to the origin (the $M^{th}$ LOR in the Fig. 1) is
\begin{equation}\label{eq3}
y=-\frac{x}{\tan\theta}-\frac{1}{2\sin\theta}
\end{equation}
The $M^{th}$ LOR intersects the upper edge of the image at $A_{0M}$ and the lower edge of the image at $A_{nM}$. This paper first calculates the length of the intersections of the LORs belonging to the calculation group with the pixels, and then determines the pixel index. Actually, they are synchronized in algorithm.

\begin{figure}[h]
\centering
\includegraphics[width=3.3in]{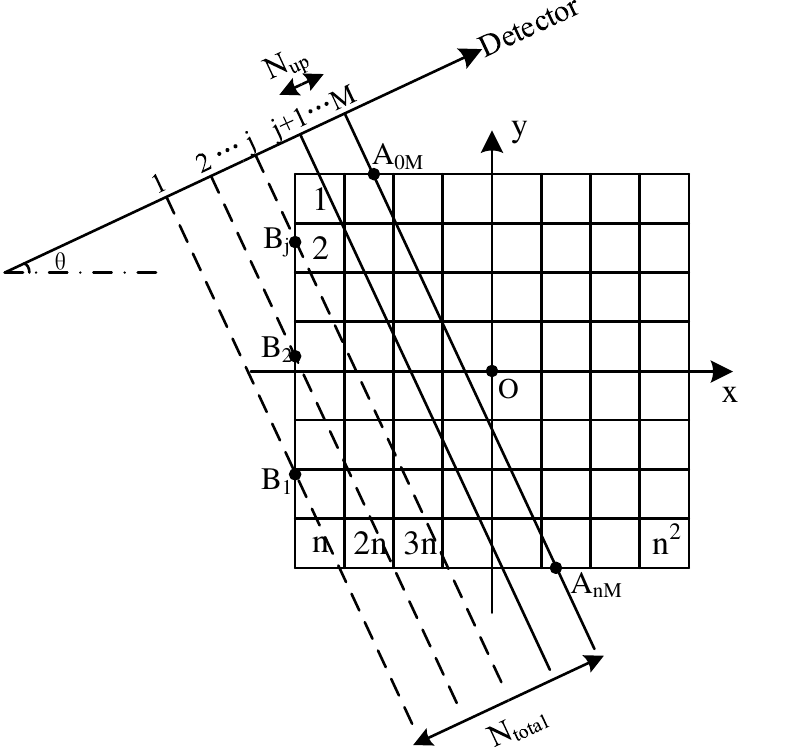}
\caption{LORs of the calculation group intersect the image of size $n \times n$.}\label{fig1}
\end{figure}

\subsubsection{Calculate the length of intersections of the LORs with the pixels}\label{sec2.2.1}
The steps are as follows:

\textbf{\uppercase\expandafter{\romannumeral1}}. Calculate the abscissa $X_{up0}$ and $X_{bottom0}$ of $A_{0M}$ and $A_{nM}$ according to formula(3). Then obtain $N_{up}$ and $N_{total}$. $N_{up}$ and $N_{total}$ are obtained by formula(4) and formula(5), respectively.

\begin{equation}\label{eq4}
N_{up}=\lfloor(\frac{n}{2}+X_{up0})\cos\theta \rfloor +1
\end{equation}

\begin{equation}\label{eq5}
N_{total}=\lfloor(\frac{n}{2}+X_{bottom0})\cos\theta \rfloor +1
\end{equation}

\textbf{\uppercase\expandafter{\romannumeral2}}.Use the isometric nature to find the abscissas of the intersections of the $N_{total}$ LORs with the upper and lower edge of the image, denoted as vectors $\bm{X_{up}}$ and $\bm{X_{bottom}}$, respectively. Intersections of the first $N_{total}-N_{up}$ LORs with line $y=\frac{n}{2}$ are beyond the image space, which are obviously meaningless, but these LORs intersect the left edge of the image at $B_i(i=1,2,\cdots,N_{total}-N_{up})$. So we correct first $N_{total}-N_{up}$ elements of $\bm{X_{up}}$ to $-\frac{n}{2}$, which is expressed as formula(6).

\begin{equation}\label{eq6}
\bm{X_{up}}=[\underbrace{-\frac{n}{2},-\frac{n}{2},\cdots,-\frac{n}{2}}_{N_{total}-N_{up}},
               \underbrace{\cdots,X_{A_{0j}},\cdots}_{N_{up}}]
\end{equation}
where $A_{0j}(N_{total}-N_{up}+1 \leq j \leq N_{total})$ is the intersection of the $j^{th}$ LOR with line $y=\frac{n}{2}$, and $X_{A_{0j}}$ is the abscissa of $A_{0j}$.

\textbf{\uppercase\expandafter{\romannumeral3}}. Find the number of pixels that passed through by the $j^{th}$ LOR, denoted as $len$. In these $N_{total}$ LORs, the first $N_{total}-N_{up}$ LORs intersect the left edge of the image, but the last $N_{up}$ LORs intersect the upper edge of the image. They are two different situations, which are discussed below in categories.
\begin{itemize}
\item [1)]
When $N_{total}-N_{up}+1 \leq j \leq N_{total}$, as shown by the solid line in Fig. 1,
$Num_j=(\lfloor \bm{X_{up}}(j) \rfloor+\frac{n}{2})\times n+1$, where $Num_j$ is the index of the first pixel that the $j^{th}$ LOR passes through.
According to the idea of SLT method, $len=n+N_Y$, where $N_Y$ is the number of intersections of the $j^{th}$ LOR with the lines parallel to the y-axis in the image space and equals to
$\lceil\bm{X_{up}}\rceil-\lfloor\bm{X_{bottom}}\rfloor+1$.
\item [2)]
When $1 \leq j \leq N_{total}-N_{up}$, as shown by the dotted line in Fig. 1,
$Num_j=\frac{n}{2}+1-\lceil \bm{Y_{left}}(j) \rceil$, where $\bm{Y_{left}}$ is the vector composed by the ordinates of the intersections at which these LORs intersect the left edge of the image. According to the idea of SLT method, $len=n+N_Y-Num_j$, where $N_Y$ has the same meaning and calculation formula as the $N_Y$ in 1).
\end{itemize}

Based on the analysis of 1) and 2), we can get the formula(7) and formula(8), where $j$ is from $1$ to $N_{total}$.
\begin{equation}\label{eq7}
Num_j=\begin{cases}
\frac{n}{2}+1-\lceil \bm{Y_{left}}(j) \rceil,1 \leq j \leq N_{total}-N_{up}\\
(\lfloor \bm{X_{up}}(j) \rfloor+\frac{n}{2})\times n+1,others
\end{cases}
\end{equation}

\begin{equation}\label{eq8}
len=\begin{cases}
n+N_Y-Num_j,1 \leq j \leq N_{total}-N_{up}\\
n+N_Y      ,others
\end{cases}
\end{equation}

\textbf{\uppercase\expandafter{\romannumeral4}}. Calculate the ordinates of the intersections of the $j^{th}$ LOR with the lines parallel to the y-axis between $x=\lceil \bm{X_{up}}(j) \rceil$ and $x=\lfloor \bm{X_{bottom}}(j) \rfloor$ in the image, denoted as $\bm{Y}$.

\textbf{\uppercase\expandafter{\romannumeral5}}. Initialize and correct the length vector $\bm{u}$. Since we assume
the LORs are only divided by the lines parallel to the x-axis, the initial value of the elements in $\bm{u}$ denoted as $d=\frac{1}{\cos\theta}$. So we initialize $\bm{u}$ to
\begin{equation}\label{eq9}
\bm{u}=[\underbrace{d,d,\cdots,d}_{len}]
\end{equation}
 In this equation, the value of $len$ can be calculated by formula(8). In fact, each intersection of the LOR with the line parallel to y-axis divides $d$ into two parts, and we call it the truncation point, as shown in $C_{0j}$ and $C_{1j}$ in Fig. 2. Therefore, each truncation point corresponds to two elements in $\bm{u}$. We use $\bm{Y}$, $\bm{X_{up}}$ and $\bm{X_{bottom}}$ to find this correspondence relationship and correct $\bm{u}$. The corrected $\bm{u}$ is the real length of the $j^{th}$ LOR passing through pixels.

\begin{figure}[h]
\centering
\includegraphics[width=2.2in]{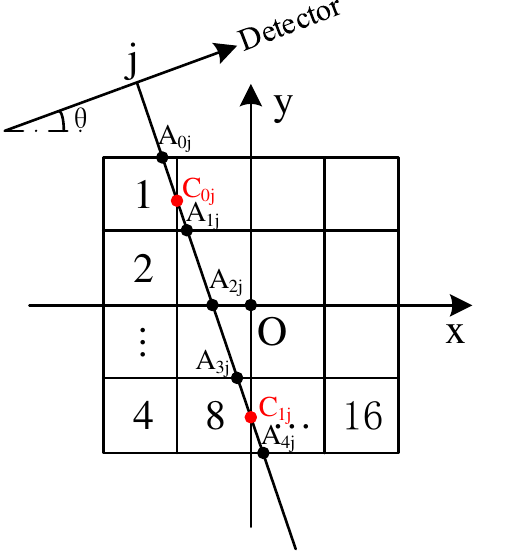}
\caption{ LORs of the calculation group intersect the image of size $4 \times 4$.
\label{fig2}}
\end{figure}

\subsubsection{Determine the pixel index}\label{sec2.2.2}
Initialize and correct the pixel index through which the $j^{th}$ LOR passes, denoted as $\bm{v}$. According to the idea of SLT method, the initial grid index adds 1 in turn. So we initialize $\bm{v}$ to
\begin{equation}\label{eq10}
\bm{v}=[\underbrace{1,1,\cdots,1}_{len}]
\end{equation}
In this equation, the value of $len$ can be calculated by formula(8). In addition, since $\bm{v}$ corresponds one-to-one with $\bm{u}$, each truncation point corresponds to two elements in $\bm{v}$. The second element is $n$ more than the first element. So the second element corresponding to the truncation point is corrected to $n$. We use the function $cumsum$ to process $\bm{v}$ , and then add $Num_j$ to $\bm{v}$. Finally we can determine the pixel index.\\

In order to understand the correction process of $\bm{u}$ and $\bm{v}$, the problem is explained by $4 \times 4$ image. As shown in Fig. 2, we take the LORs intersecting the upper and lower edges of the image as an example. The LOR intersecting the left and lower edges of the image is similar. $A_{0j}, A_{1j}, A_{2j}$ and $A_{3j}$ are the intersections of the $j^{th}$ LOR with the lines parallel to the x-axis in the image. $C_{0j}$ and $C_{1j}$ are the intersections of the $j^{th}$ LOR with the lines parallel to the y-axis.

From Fig. 2 we can get: $N_Y=2$, $len=6$, $Num_j=1$. And the $j^{th}$ LOR is divided into four segments by the straight line parallel to the x-axis in the image, so $\bm{u}$ and  $\bm{v}$ are divided into four groups. But two groups of them are divided by the lines parallel to the y-axis. So we initialize $\bm{u}=[\underbrace{d,d,\cdots,d}_{6}]$ and $\bm{v}=[\underbrace{1,1,\cdots,1}_{6}]$ and correct them. The correction process is shown in Fig. 3.

\begin{figure}[h]
\centering
\includegraphics[width=3.4in]{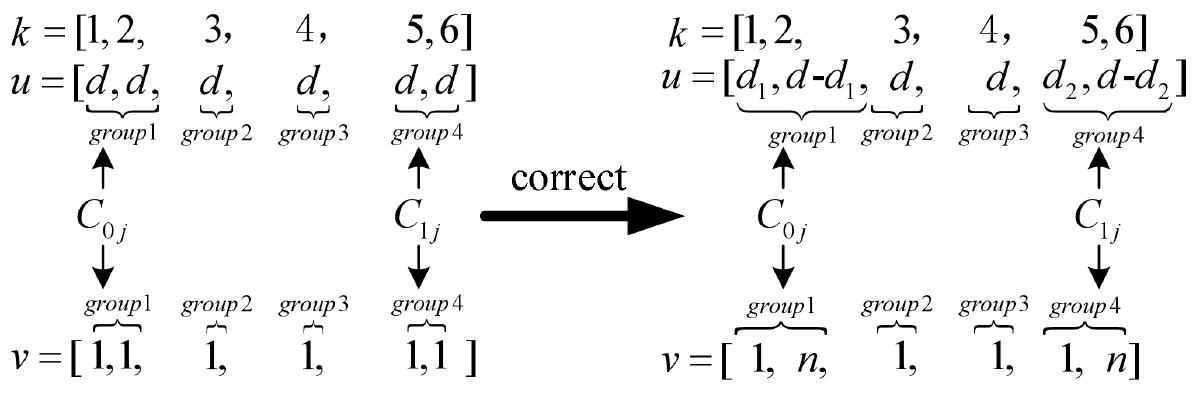}
\caption{The correction process of the length vector $\bm{u}$ and the grid index vector $\bm{v}$ for the case of Fig. 2, wherein $d_1=\left|A_{0j}C_{0j} \right|$, $d_2=\left|A_{3j}C_{1j} \right|$.
\label{fig3}}
\end{figure}


\section{Experiments And Results}\label{sec3}
Through the use of MATLAB 2016b environment, we carried out our experiments on a PC configured for Intel Core i3-6006U CPU@2.00GHz. Experiment 1 used the SLT algorithm to calculate the projection, and evaluated the projection data at $30^\circ$. Experiment 2 compared the SLT algorithm with the traditional Ray-Tracing algorithm in the performance of the projection on time and space. Experiment 3 separately projected the image at different angles to verify the advantages of our algorithm in the speed of small angles projection. Experiment 4 used FBP and different iterative reconstruction algorithms (ART Reconstruction and MLEM(Maximum Likelihood Expectation Maximization) Reconstruction) to reconstruct the Shepp-Logan of different sizes, and tested the performance of the SRM calculated by SLT algorithm.\\
\textbf{Experiment 1:} The prosthesis is set to a circle (circular density is 1) with a diameter of 64, 128 and 256, respectively. The standard line integral of the circle can be calculated directly. For a circle with a density of 1, its line integral formula is as equation(11), where $R$ is the radius of the circle, $t$ is the intercept, $P$ is the integral value corresponding to $t$. The projection performance of this algorithm is evaluated by comparing with $P(t)$ and Radon algorithm.
\begin{equation}\label{eq11}
P(t)=2\sqrt{R^2-t^2}
\end{equation}


\begin{figure}[b]
\centering{\includegraphics[width=3.4in]{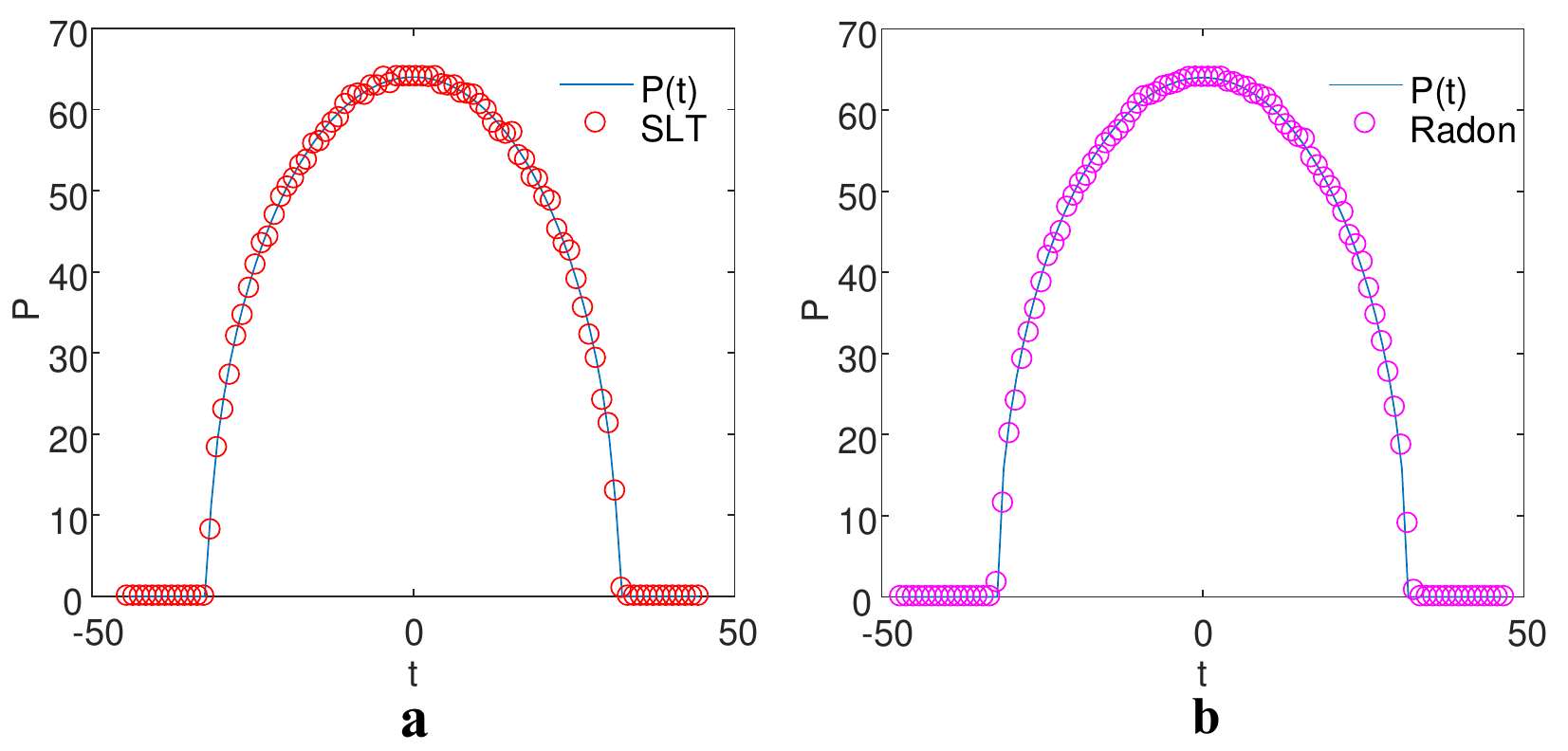}}
\caption{Comparing the projection data of the circle with a radius of 64 calculated by SLT algorithm and Radon algorithm. $P$ is the projection data. $t$ is the intercept.
\figfooter{a}{Projection values of SLT}
\figfooter{b}{Projection values of Radon}\label{fig4}}
\end{figure}

\begin{figure}[b]
\centering{\includegraphics[width=3.4in]{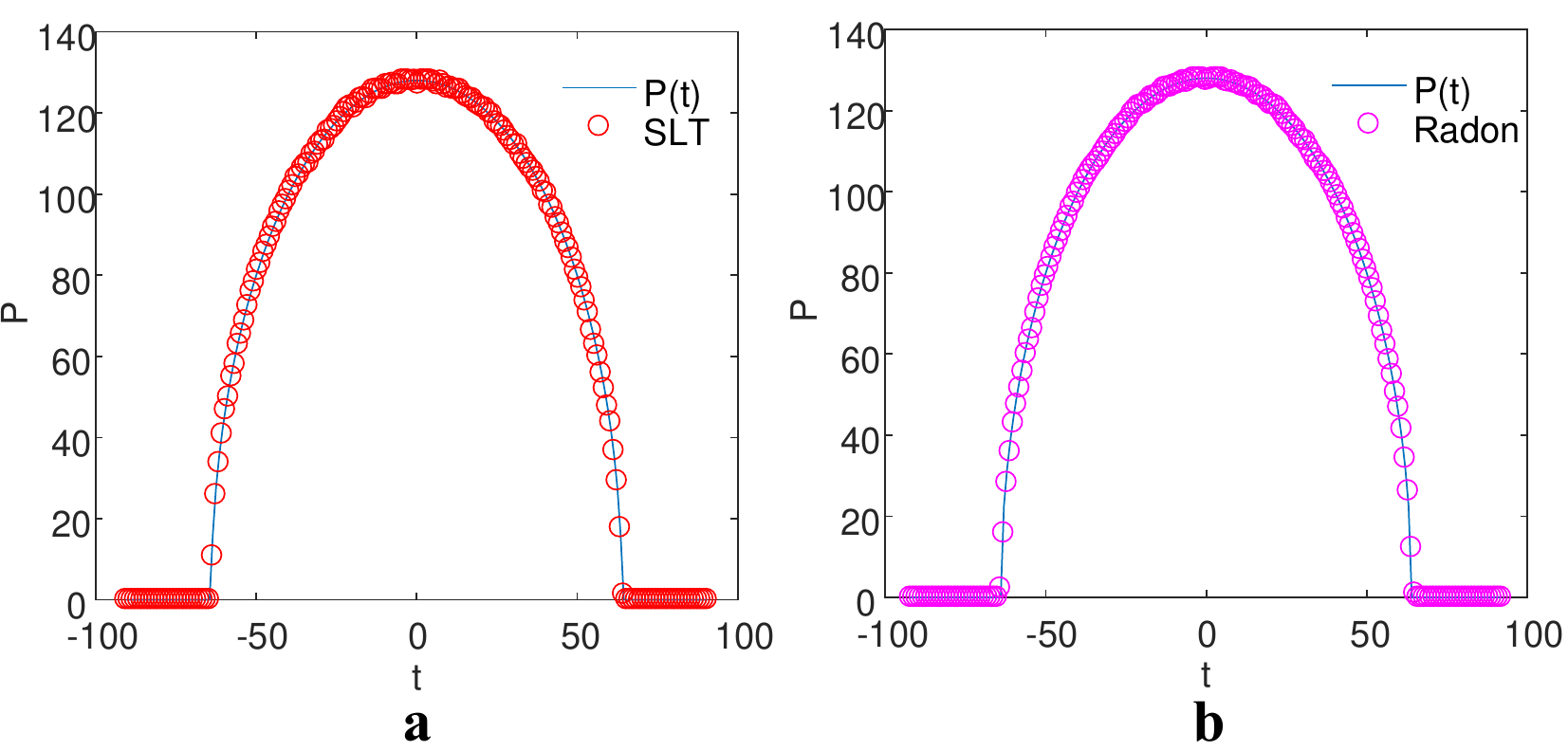}}
\caption{Comparing the projection data of the circle with a radius of 128 calculated by SLT algorithm and Radon algorithm. $P$ is the projection data. $t$ is the intercept.
\figfooter{a}{Projection values of SLT}
\figfooter{b}{Projection values of Radon}\label{fig5}}
\end{figure}

\begin{figure}[b]
\centering{\includegraphics[width=3.4in]{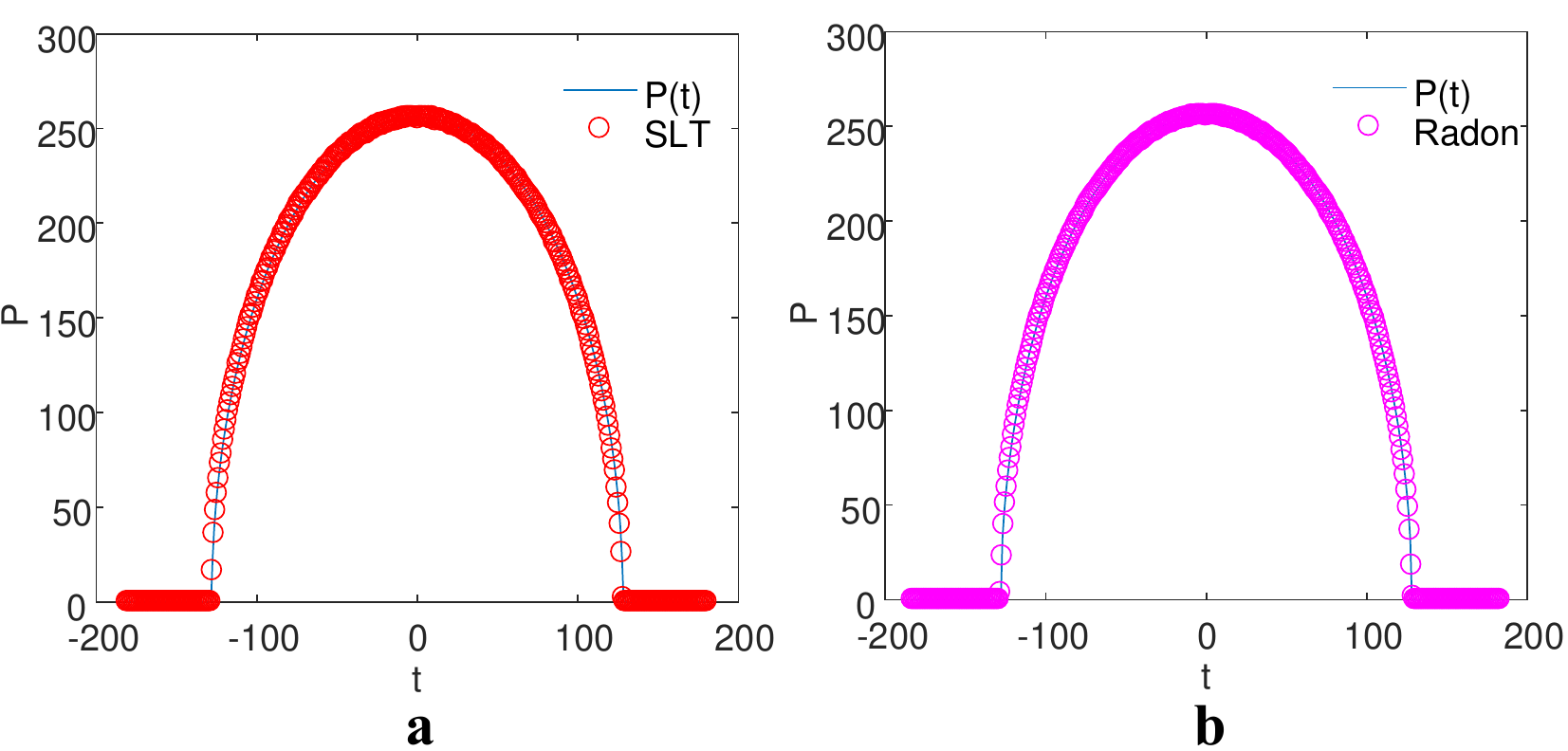}}
\caption{Comparing the projection data of the circle with a radius of 256 calculated by SLT algorithm and Radon algorithm. $P$ is the projection data. $t$ is the intercept.
\figfooter{a}{Projection values of SLT}
\figfooter{b}{Projection values of Radon}\label{fig6}}
\end{figure}




The experimental results are shown in Fig. 4, Fig. 5 and Fig. 6. As can be seen from the Fig. 4, Fig. 5 and Fig. 6, for the circles of the same size,  the forward projection data calculated by SLT algorithm in this paper is very close to that calculated by Radon algorithm and the standard line integral value. It means that the SLT algorithm in this paper does not have many errors in calculating the numerical value.\\
\textbf{Experiment 2:} This experiment uses the SLT algorithm and the traditional Ray-Tracing algorithm to project the Shepp-Logan of $64 \times 64$, $128 \times 128$ and $256 \times 256$ two hundred times, and evaluate the performance of the SLT algorithm on both time and space.

The numerical results are given in table 1. As can be seen from the table, when the image size is $64 \times 64$, The SLT algorithm in this paper calculates projection more than 6 times faster than the traditional Ray-Tracing algorithm, and consumes less memory than the traditional Ray-Tracing algorithm. It can be imagined that when the image size doubles, in order to calculate the length vector, the intersections of LORs with lines parallel to x-axis or y-axis calculated by traditional methods are about four times as many as the original ones, so the time spent is about four times. The truncation points are about twice as many as the original ones, so it takes about two times. Therefore, this advantage will become more obvious as the size of the image becomes larger. It proves that the SLT algorithm has excellent performance in calculating projection.
\begin{table}[h]
\processtable{The time and space consumed by the projection with SLT algorithm and the traditional Ray-Tracing algorithm\label{tab1}}
{\begin{tabular}{@{\extracolsep{\fill}}l!{\vrule width1pt}llll}\Xhline{1pt}
Projection \\algorithm&Image size & $64 \times 64$ &$128 \times 128$&$256 \times 256$\\\Xhline{0.5pt}
\multirow{2}{*}{SLT}&Mean time(s) & 0.0684 &  0.1743  &  0.5384\\
{}&Space(bytes) & 302200  & 672792  & 1598968\\\Xhline{0.5pt}
\multirow{2}{*}{Ray-Tracing}&Mean time(s) &0.4395  & 1.6978   &  8.3611\\
{}&Space(bytes) &408008  & 1072200  & 3192776 \\\Xhline{1pt}
\end{tabular}}{}
\end{table}
\\
\textbf{Experiment 3:} In this experiment, the SLT algorithm and the traditional Ray-Tracing algorithm are used to project the Shepp-Logan of $256 \times 256$ at different angles two hundred times respectively, and the time consumed is compared. The results are shown in Table 2 and Fig. 7.

\begin{table}[h]
\processtable{The time consumed by the projection with SLT algorithm and the traditional Ray-Tracing algorithm at different angles\label{tab2}}
{\begin{tabular}{@{\extracolsep{\fill}}l!{\vrule width1pt}llll}\Xhline{1pt}
Projection&Projection&{}&{}&\\
algorithm&angle&$1^\circ - 5^\circ$ &$21^\circ - 25^\circ$&$41^\circ - 45^\circ$\\\Xhline{0.5pt}
SLT&Mean time(s) & 0.0093 &   0.0145  &   0.0180\\
Ray-Tracing&Mean time(s) &0.2141  &   0.2290   &   0.2369\\\Xhline{1pt}
\end{tabular}}{}
\end{table}

\begin{figure}[h]
\centering
\includegraphics[width=3.2in]{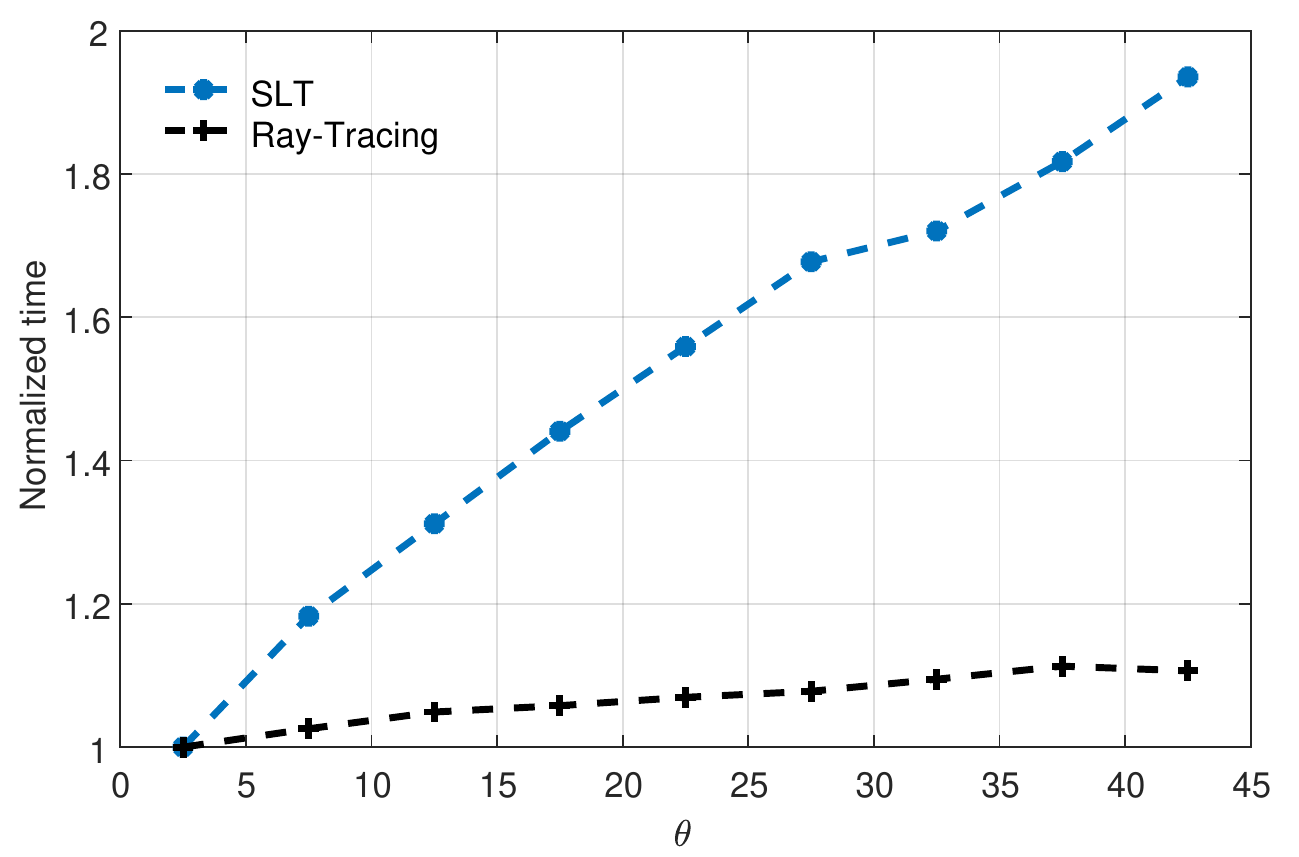}
\caption{\emph{Time graph of forward projection at different angles using the SLT algorithm and the conventional Ray-Tracing algorithm. Note that the normalized time is a multiple of the time at $1^\circ - 5^\circ$ of corresponding algorithm.}
\label{fig7}}
\end{figure}

As can be seen from Table 2 and Fig. 7. The traditional Ray-Tracing algorithm does not have much difference in the time of projection at different angles. For the SLT algorithm in this paper, as the angle decreases, the less time it takes, and the time at $1^\circ \leq \theta \leq 5^\circ$ is about half at $41^\circ \leq \theta \leq 45^\circ$. This is because small angle projection has fewer truncation points and it leads faster correction speed. It can be considered that the SLT algorithm in this paper has obvious advantages in small angle projection.\\
\textbf{Experiment 4:} In order to verify the feasibility of SLT algorithm, this experiment performs image reconstruction with the SRM calculated by SLT algorithm. Reconstructed images are Shepp-Logan of $128 \times 128$, $256 \times 256$, $512 \times 512$, and this experiment uses ART and MLEM to iterate 100 times, the result is shown in Fig. 8. In addition, in order to evaluate the performance of the algorithm better, this experiment uses FBP for 100 times repeated reconstructions and adds $64 \times 64$ Shepp-Logan reconstruction. Their numerical results of the reconstruction are shown in Table 3 and Fig. 9.

\begin{figure}[h]
\centering{\includegraphics[width=3.3in]{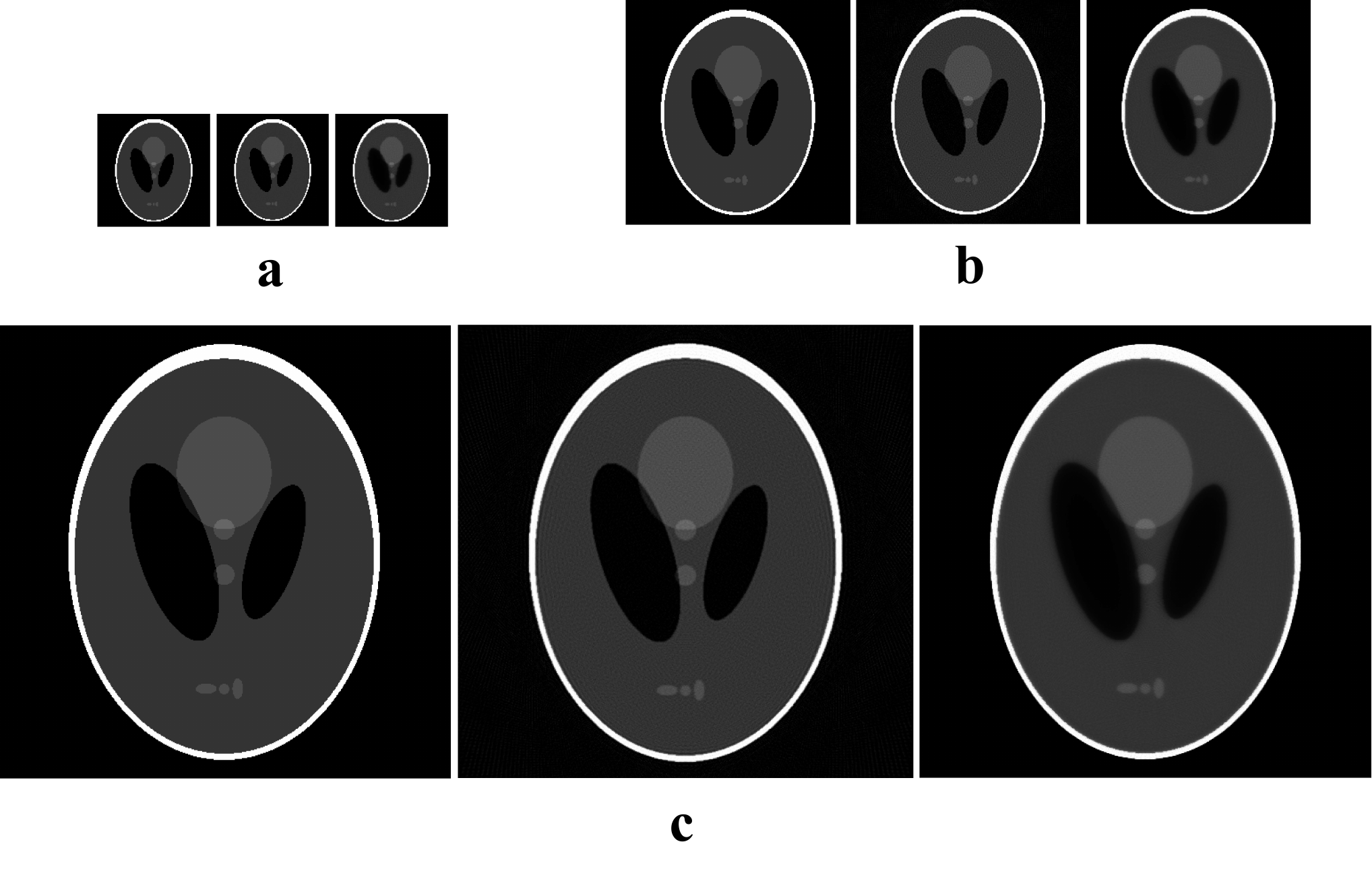}}
\caption{Reconstruction of Shepp-Logan of different sizes. In each subfigure, the order from left to right is the original image, reconstructed image by ART algorithm and reconstructed image by MLEM algorithm.
\figfooter{a}{Reconstruction of Shepp-Logan of size $128 \times 128$}
\figfooter{b}{Reconstruction of Shepp-Logan of size $256 \times 256$}
\figfooter{c}{Reconstruction of Shepp-Logan of size $512 \times 512$}\label{fig8}}
\end{figure}

\begin{table}[h]
\processtable{The time it takes to reconstruct Shepp-Logan of different sizes using FBP, ART and MLEM algorithms\label{tab1}}
{\begin{tabular*}{20pc}{@{\extracolsep{\fill}}cccc@{}}\toprule
Image  &FBP(s)  & ART(s) &MLEM(s)\\
size  & & & \\
\midrule
$64 \times 64$  &9.1283  &44.4418 &52.0905\\
$128 \times 128$  &20.9536  &105.0475 &119.2540\\
$256 \times 256$  &70.4952  &240.1283 &267.6808\\
$512 \times 512$  &238.1642  &551.2434 &598.7615\\
\botrule
\end{tabular*}}{}
\end{table}

\begin{figure}[h]
\centering{\includegraphics[width=3.5in]{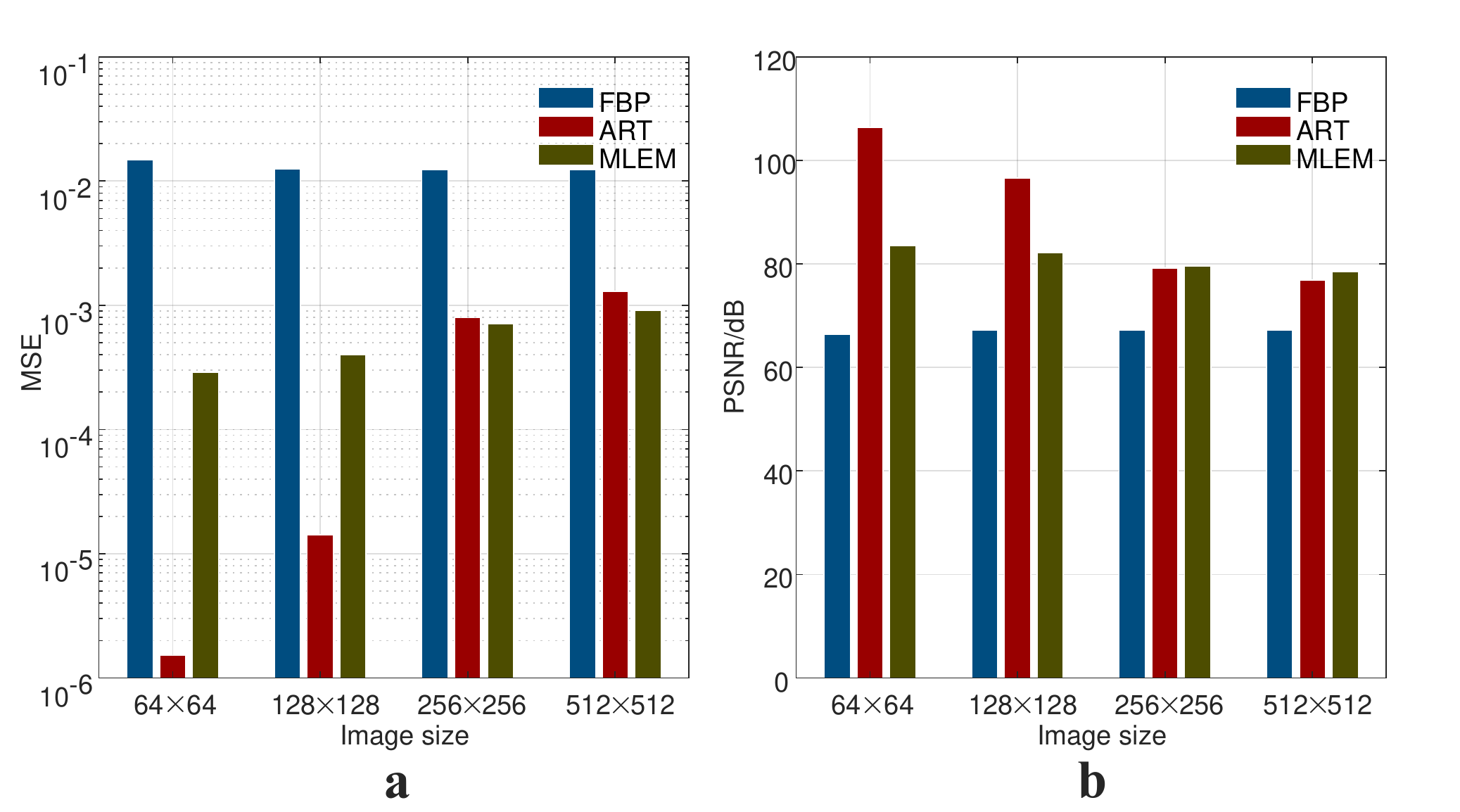}}
\caption{The comparison of performance among FBP, ART and MLEM.
\figfooter{a}{The comparison of MSE(Mean Square Error) among FBP, ART and MLEM}
\figfooter{b}{The comparison of PSNR(Peak Signal to Noise Ratio) among FBP, ART and MLEM}\label{fig9}}
\end{figure}

As shown in Table 3, when reconstructing Shepp-Logan of different sizes, FBP takes the minimum time and MLEM takes the maximum time. Fig. 9 shows that the accuracy of FBP is always worst. When the Shepp-Logan size is $64 \times 64$, the ART algorithm is superior than the MLEM algorithm by an order of magnitude in accuracy. However, with the increase of the Shepp-Logan size, the variance of the ART algorithm is larger, and the SNR is reduced accordingly. But the variance and SNR of the MLEM algorithm change less, so the accuracy of the MLEM algorithm first approaches and then even exceeds the ART algorithm. According to the above analysis, we know FBP has high real-time capability but low precision. ART and MELM are high precision but time-consuming. It can also be seen that ART and MLEM have their specific advantages and disadvantages on the reconstruction based on SLT. Those above results prove that the SLT algorithm of this paper has certain significance for image reconstruction.

At the same time, SRM is a large sparse matrix, so the SRM of SLT algorithm only stores non-zero values, which greatly reducing the storage space. For the Shepp-Logan image of size $128 \times 128$, the original SRM needs to consume $128 \times 128 \times 180 \times 182 \times 8=4.2939\times 10^{9}$ bytes of memory space, and the SRM of SLT algorithm needs to consume $66863392$ bytes of memory space(In fact, there are three variables stored. $x$ stores the row of non-zero elements, $y$ stores the column of SRM non-zero elements , $val$ stores the value of non-zero elements). The storage space of SRM calculated by SLT algorithm is about one sixty-fourth of that calculated by the traditional Ray-Tracing algorithm.

\section{Discussion And Conclusion}\label{sec4}
In this work, we have proposed SLT method in combination with the idea of symmetric blocks. Compared with the traditional Ray-Tracing methods, the SLT method avoids calculating the redundant intersection coordinates of LORs with pixels in image space. It aims at improving the reconstruction speed by using truncation points to correct the length vector and the grid index vector. So the reconstruction time required is shorter. Simultaneously, due to fewer truncation points at small angles, the correction process takes less time, which is favorable for the acceleration of small-angle image reconstruction. Since only the non-zero values of the SRM are calculated and stored, the storage space is reduced while ensuring the accuracy, and the speed of SLT algorithm is further improved. Therefore, the proposed method is suitable for iterative reconstruction and infinitesimal analysis method \cite{27}.

In order to verify the superiority of our algorithm, we conducted four numerical experiments in MATLAB. By comparing projection data at 30 degrees of SLT algorithm with that of Radon algorithm, we proved the feasibility of SLT algorithm. In addition, we used SLT algorithm and traditional Ray-Tracing algorithm to project different sizes of Shepp-Logan, and concluded that the SLT algorithm has obvious advantages in speed and space complexity. At the same time, we compared the time consumed by SLT algorithm and traditional Ray-Tracing algorithm at different angles, and found that SLT algorithm has a more obvious time advantage at small angles, which is beneficial to accelerate the process of image reconstruction at small angles. Finally, we applied different back projection algorithms to verify the SRM calculated by SLT algorithm. The results proved that the matrix can be used in iterative reconstruction process.

Our research is built on a discretization process, which is a common approach in image reconstruction \cite{25}. And it is embodied both in the pixelation of the image and the idealization of LORs. This means discrete projection is different from actual ray energy attenuation along the path. Owing to the existence of the noise, deviations caused by it will accumulate in image reconstruction, and finally affect the quality of the reconstructed image \cite{28}. What's more, although SLT algorithm has high speed at small angles, the speed of large angle reconstruction is relatively slow, which becomes the crucial factor to limit the reconstruction speed.

Therefore, in the practical application, the solution proposed in this paper still needs to be further perfected in order to strengthen the resistance to noise. Besides, solving the problem of large-angle reconstruction speed will be the future research direction of this algorithm, which will be of great significance for the development of iterative methods and fast image reconstruction. Last but not least, the SLT algorithm can also be applied to 3D PET image reconstruction and it will perform better.

\section{Acknowledgments}\label{sec5}
The authors would like to thank their colleague Anyi Li for his help in experiments, and they would also like to express their gratitude to PET Lab of Nanchang University for its academic support.


\end{document}